\documentclass[prd,aps,floatfix,nofootinbib,preprint ,tightenlines ]{revtex4}
\usepackage{dcolumn}
\usepackage{amsmath}
\usepackage{latexsym}
\usepackage{graphicx}
\usepackage{bm}

\def\svev#1{\left\langle #1\right\rangle}       

\long \def \blockcomment #1\endcomment{}

\newcommand{\bee}{\begin{equation}}
\newcommand{\ee}{\end{equation}}
\newcommand{\beea}{\begin{eqnarray}}
\newcommand{\eea}{\end{eqnarray}}

\begin{document}
\title{%
Lattice studies of QCD-like theories with many fermionic degrees of freedom}
\author{Thomas DeGrand}%
\affiliation{Department of Physics,
University of Colorado, Boulder, CO 80309, USA}

\begin{abstract}
I give an elementary introduction to the study of gauge theories coupled
to fermions with many degrees of freedom. Besides their intrinsic interest,
these theories are candidates for nonperturbative extensions of the Higgs sector of
 the Standard Model.
 While related to QCD,
these systems can exhibit very different behavior from it: they
can possess 
a running gauge coupling with an infrared attractive fixed point (IRFP).
I briefly survey recent lattice work in this area.
\end{abstract}

\maketitle

\section{Introduction}
In the Standard Model of elementary particle physics, the electroweak 
 coupling constants are small and much of the phenomenology
is perturbative. The strong interactions are nonperturbative: quarks are permanently confined.
But what about the unseen sector of the Standard Model, which is responsible for
electroweak symmetry breaking. Is it weakly interacting, described by one or more fundamental
Higgs fields, or is it also strongly interacting? The textbook Standard Model
assumes it is the former. However, the idea that
 there is new strongly interacting dynamics in the Higgs sector is long lived.
The main motivation for this is the hierarchy or ``naturalness''
or ``fine-tuning'' problem: the mass of the Higgs (or of any
fundamental scalar field) depends quadratically on the ultraviolet cutoff
$\Lambda$,
mass at the cutoff scale $m_0$ and  $\lambda$ the bare Higgs self-coupling,
\bee
m_H^2 = m_0^2 + \frac{3}{4\pi}\lambda \Lambda^2
\label{eq:hier}
\ee
To achieve a Higgs mass which is much (much) less than $\Lambda$ requires a delicate
cancellation among the various bare quantities. If the Higgs is composite,
the fine tuning problem is softened. For example, if the Higgs is a bound state
of some new fermions, the quadratic dependence on $\Lambda$ is transformed (at
one loop) into
 $M_H \sim \Lambda \exp(-c/g^2(\Lambda))$, or a value
near the scale where the dynamics which forms the Higgs becomes strong.

The dynamics which gives the $W$ and $Z$ their masses is that they ``eat'' Goldstone
bosons which were present in the ungauged theory. Any kind of Goldstone can be eaten,
and the pionic analog of the new fermionic bound states makes a tasty meal.
In the Lagrangian, the vacuum expectation value $v$ of the fundamental scalar field is replaced by
the pseudoscalar decay constant of the bound state, $F_\pi$, which must be set (by hand) to its
real world value of $v \sim F_\pi \sim 250$ GeV.
This particular idea is called ``technicolor'' \cite{Weinberg:1975gm,Susskind:1978ms}.
(For a review, see \cite{Hill:2002ap}.
A recent introduction to the (continuum) theoretical background is Ref.~\cite{Contino:2010rs}.)

Lattice techniques have been very successful at dealing with the nonperturbative dynamics
of QCD and indeed the lattice is now the source of many high precision calculations
of strong interaction masses and matrix elements. It is natural to think that lattice
techniques can be applied to strongly interacting dynamics which may appear at the energy
of the Large Hadron Collider or beyond.

Before going to the lattice, let us think analytically.
Suppose we have an
SU($N_c$) gauge theory with $N_f$ flavors of fermions of mass $m$, in representation $R$.
 The Gaussian
fixed point at $g^{2}=0$, $m=0$ is well understood perturbatively.
The mass is a relevant operator and it will presumably  remain
so even at strong gauge coupling. $m=0$ is a critical surface and
we want to investigate the running of the gauge coupling along it.
The analysis can begin in perturbation theory\cite{Caswell:1974gg}.
The gauge coupling's beta function is
\bee
\beta(g^2)=\frac{dg^2}{d\log q^2}=-\frac{b_1}{16\pi^2}g^4-\frac{b_2}{(16\pi^2)^2}g^6+\cdots,
\label{2loopbeta}
\ee
where
\beea
b_1&=&  \frac{11}{3}\, C_2(G) - \frac{4}{3}\,N_f T(R) \label{2loopbeta1}\\
b_2&=& \frac{34}{3}\, [C_2(G)]^2
  -N_f T(R) \left[\frac{20}{3}\, C_2(G) 
  + 4 C_2(R) \right].
\label{2loopbeta2}
\eea
Here $C_2(R)$ is the value of the quadratic Casimir operator in representation $R$ 
($G$ denotes the adjoint representation, so $C_2(G)=N_c$), while $T(R)$ is the
 conventional trace normalization.

\begin{figure}
\begin{center}
\includegraphics[width=0.7\textwidth,clip]{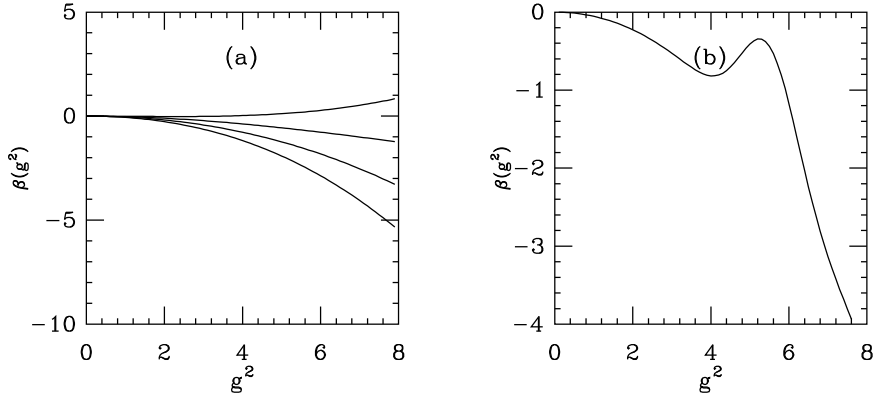}
\end{center}
\caption{
Examples of beta functions. Panel (a) shows the two-loop beta function for $SU(3)$ 
gauge theory with
(from the bottom) $N_f=2$, 6, 10, 14 flavors of fundamental representation fermions.
The $N_f=14$ beta function starts off negative before turning positive.
If $N_f>16.5$ the beta function is everywhere positive.
Panel (b) shows the ``walking technicolor dream'' beta function, under which the coupling
constant would run out quickly to about $g^2=5.5$, then runs slowly before taking
off again. }
\label{fig:bfnex}
\end{figure}

These theories have an ultraviolet-attractive or infrared-repulsive or Gaussian fixed point
at $g^2=0$.
 In perturbation theory, three things can happen as we flow to the infrared:
\begin{itemize}
\item $b_1 < 0$: the gauge coupling runs to the Gaussian fixed point in the infrared.
Theories that do this are ``trivial.''
\item $b_1 >0$, $b_2>0$: the gauge coupling runs to a large value in the infrared, and
long distance dynamics is nonperturbative. This is the situation
 for ordinary QCD ($N_c=3$ with $N_f=2$ or 3). In that case, the theory is confining
 and chiral symmetry is broken spontaneously. 
\item $b_1 >0$, $b_2<0$: In this case it is possible that there is a critical coupling $g^2_*$
where $\beta(g^2_*)= 0$. The coupling runs to this value and sticks there.
One says that there is an ``infrared attractive fixed point'' or IRFP.
The range of values of $N_c$ and $N_f$ for which this occurs is called the
 ``conformal window.'' 
Some examples of beta functions are shown in Fig.~\ref{fig:bfnex}.
\end{itemize}

Conventional strong-coupling models for the Higgs system require theories which
 make the confinement --
chirally broken choice. They have a set of Goldstone bosons which are fermion bound states,
and which are food to be eaten by the electroweak gauge bosons. The Higgs particle
is the analog of the sigma meson, a mesonic bound state with scalar quantum numbers.
There are also towers of bound states which either have to be explained away if they are light enough
to have already been discovered, or are potential ``new physics'' for the Large Hadron Collider.

Theories with an IRFP are like condensed matter systems possessing
second order phase transitions, at their critical points. They do not have a particle spectrum:
all correlation functions decay algebraically at long distances.
(Georgi \cite{Georgi:2007si} coined the term ``unparticles'' to describe 
the excitations to particle physicists.) For the most part, these theories are
not taken to be candidates for beyond - Standard Model physics
but they do occupy a leading role in theoretical particle physics: they
are the CFT (conformal field theories) of AdS-CFT which are widely used as avatars for analog
calculations of strong-interaction dynamics.

Not just any confining theory is a candidate for beyond - Standard Model physics.
In addition to generating gauge boson masses, the Higgs sector is responsible for
the masses of the ordinary fermions. Technicolor was invented to give 
the gauge bosons their masses. To generate fermion masses, one needs
 some new dynamics (``extended technicolor'') at some even higher scale.
It is a delicate business to simultaneously generate acceptable fermion masses
while avoiding too-large flavor changing neutral currents.
Successful phenomenology appears to require ``walking:'' which (to quote T.~Appelquist
 \cite{appel}) is   ``a
theory that is outside the conformal window, but close to it, with the
would-be IRFP somewhat
super-critical,'' so that the running coupling evolves very slowly over a wide range of scales.

In addition, phenomenology makes a special demand on the running of the mass parameter.
The observable of interest
 is the anomalous dimension $\gamma_m$ of the mass operator $\bar\psi\psi$.
\bee
\mu \frac{\partial m(\mu)}{\partial \mu} = -\gamma_m(g^2) m(\mu).
\ee
For future reference, in lowest order in perturbation theory,
\bee
\gamma_m= \frac{6 C_2(R)}{16\pi^2} g^2.
\label{eq:gmpt}
\ee
Successful technicolor models typically need $\gamma_m \sim 1$ while the coupling walks
 \cite{Chivukula:2010tn}.
Scenarios for how this may be achieved typically depend on a running gauge coupling like the one
shown in Fig.~\ref{fig:bfnex}b. Starting at some extremely high scale where the coupling is
 small,
it runs quickly out to some intermediate low scale, where the curvature of the
 beta function up to 
near zero causes the coupling to stall.  At the same time $\gamma_m$
is supposed to run out to a large, nonperturbative value.

New physics at a high energy scale can influence low energy observations
by inducing extra operators in the low energy effective theory. These new operators
can, in turn, affect precision electroweak measurements \cite{Skiba:2010xn}. 
It could happen, that one's proposed new dynamics might already be ruled out by them.

If the system is conformal at zero fermion mass $m$, then near $m=0$ the
correlation length $\xi$ scales as
\bee
\xi \sim m^{-{1}/{y_m}}
\label{eq:xim}
\ee
where $y_m = 1+\gamma_m(g_*^2)$ is the leading relevant exponent
 of the system (in the language of critical phenomena).

Unitarity bounds  for
 conformal field theories\cite{Mack:1976pa,Grinstein:2008qk} constrain the scaling dimension of
the leading scalar operator
to lie in the range $3>d=4-y_m>1$.
For our technicolor candidates, this operator is $\svev{ \bar \psi \psi}$, the ``techni-''
condensate, nonzero because a nonzero fermion mass explicitly breaks chiral symmetry.
At the top of the conformal window, $y_m=1$ its free field value. For large $N_c$
and $N_f$ it is possible (see Ref.~\cite{Grinstein:2008qk})
 to tune the $N_f/N_c$ ratio so that $g_*$ is order $\epsilon$
and then $\gamma_m\sim \epsilon$ and $y_m=1+O(\epsilon)$. It is believed
 that $d$ decreases,
or $\gamma_m$ increases, as one moves to the bottom of the conformal window.

Nothing is known about how the conformal window closes, or more mundanely,
what happens at the conformal -- chirally broken and confining boundary.
The only way I understand this question is by imagining some continuously variable parameter
which can carry one across the boundary. The value of $N_f/N_c$ at large $N_c$ may
be such a parameter.
One possibility (which is what happens in Fig.~\ref{fig:bfnex}(a)) is that a zero
of the beta function  ``walks in'' from infinite $g^2$. Another possibility,
inspired by Fig.~\ref{fig:bfnex}(a), is that a single new fixed point just appears
on the real axis, coming out of the complex $g^2$ plane, and then splits into
an IRFP (the conformal fixed point) and a second UV fixed point at larger $g^2$ \cite{Kaplan:2009kr}.
There is also an extensive literature relating the lower end of the conformal window
to large $y_m$, with $y_m=2$ perhaps having a connection with physics which
closes the window.
 (See
Refs.~\cite{Cohen:1988sq,Braun:2006jd,Appelquist:1996dq,Appelquist:1998rb,gardi}.)
To me, this connection is Eq.~\ref{eq:xim}, which, when the correlation length is
replaced by $1/M$ for some (the pseudoscalar?) mass, becomes
\bee
M^{y_m} \propto m_q.
\ee
The value $y_m=2$ is (coincidentally?) the Gell-Mann, Oakes, Renner relation relevant to
the additional explicit breaking (by the fermion mass) of spontaneously broken chiral symmetry.

Thus, a rich program presents itself to anyone who wants to look for nonperturbative
beyond - Standard Model physics:
\begin{itemize}
\item Where does any particular theory sit, inside or outside the conformal window?
\item What is the value of $\gamma_m$, either as a function of $g^2$ or at any fixed point?
\item If the theory is confining, what is its particle spectrum? What is the Higgs
 mass and what are its couplings?
\item Does the theory survive precision electroweak constraints, or predict observable
deviations from pure Standard Model predictions?
\end{itemize}

\begin{figure}
\begin{center}
\includegraphics[width=0.5\textwidth,clip]{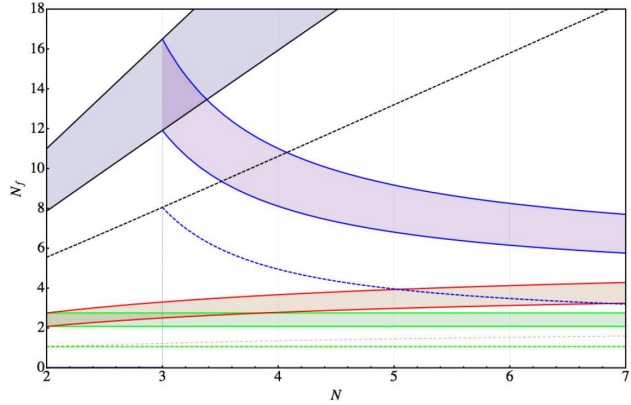}
\end{center}
\caption{
Conjectured phase diagram (due to Ref.~{\protect{\cite{Dietrich:2006cm}}})
 for nonsupersymmetric theories with fermions
in the: i) fundamental representation (blue), ii) two-index
antisymmetric representation (purple), iii) two-index symmetric
representation (red), iv) adjoint representation (green) as a
function of the number of flavors $N_f$ and the number of colors $N$. The
shaded areas depict the corresponding conformal windows from their calculation.
The dashed curve represents the change of sign in the second
coefficient of the beta function.  }
\label{fig:ds}
\end{figure}

Already the first point is challenging. Fig.~\ref{fig:ds} shows a phase diagram from an
approximate analytic calculation from Ref.~\cite{Dietrich:2006cm}. This figure has served
as the target for many lattice calculations.

\section{Beyond perturbation theory}
We do not have to restrict ourselves to a perturbative analysis of these
theories. instead, let us think about them as we would any critical
statistical system, using the language of fixed points, critical surfaces,
and renormalization group flow.
(For a related discussion, see Ref.~\cite{DeGrand:2009mt}.)

\begin{figure}
\begin{center}
\includegraphics[width=0.5\textwidth,clip]{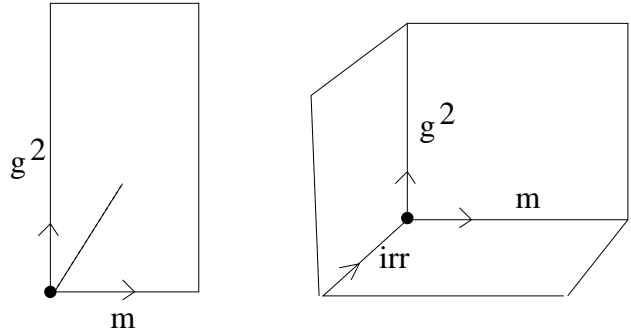}
\end{center}
\caption{Schematic coupling constant space of a QCD-like theory. The left panel shows only the relevant mass and marginally
relevant gauge couplings, with IR flow lines and a renormalized trajectory. The right panel
adds a single irrelevant direction.}
\label{fig:fullqcd}
\end{figure}

A theory like QCD, labeled with a gauge coupling constant $g^2$ and a single fermion mass $m$,
 has a Gaussian fixed point at $(g^2,m)=(0,0)$ (see  the left panel of Fig.~\ref{fig:fullqcd}). Under
a sequence of real space blocking transformations (flows to the IR, in the language
of the last section) both the running coupling and the mass will grow. To show that flow,
I have drawn arrows in the figure. Of course, the most general theory which one can write down
at the cutoff scale will also contain irrelevant couplings. The critical surface
is the surface of $(g^2,m)=(0,0)$ projected onto the space of all bare couplings, and under
blocking to the IR, flows beginning on the critical surface will carry the coupling to 
some critical point on the critical surface. Off the critical surface, flows carry us ever farther
away from it.

\begin{figure}
\begin{center}
\includegraphics[width=0.5\textwidth,clip]{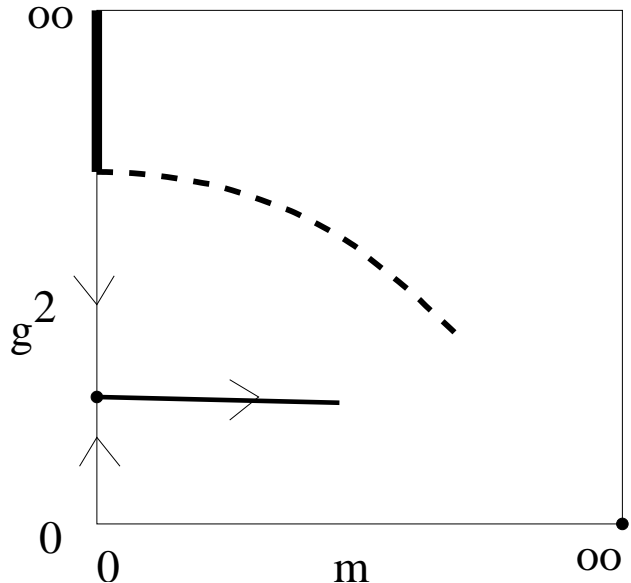}
\end{center}
\caption{A schematic picture of the parameter space of a theory with an IRFP. The part of the $m=0$
axis drawn with a thin line is on the critical surface containing the IRFP. I have also put a dot
at $(g^2,m)=(0,\infty)$ for the pure gauge theory. The thick lines show possible
strong coupling crossover or transition lines.
The line emanating from the IRFP is a schematic renormalized trajectory.}
\label{fig:irfp2}
\end{figure}

The analog of the preceding figure for the case of a theory with an IRFP is shown in Fig.~\ref{fig:irfp2}.
Again the mass is relevant. However, this time the gauge coupling flows to an IRFP, meaning that
the gauge coupling itself is irrelevant (more precisely, $g^2-g_*^2$ is an irrelevant direction).
The critical surface includes part of the $g^2$ axis. 
Precisely at $m=0$ the theory is critical; all correlation functions decay
algebraically.

Since the mass is relevant, RG flow
carries us off the $m=0$ axis. Since it is the relevant perturbation, it controls the correlation
length through Eq.~\ref{eq:xim}.
 It is easiest
to express the correction to the free energy ($D$ is the dimension, $u_i$ is any irrelevant operator),
\beea
f_s(m,u_i) &=& m^{D/y_m} f_s( m_0,u_{i0} +
(u_i-u_{i0})\left(\frac{m}{m_0}\right)^{|y_i|/y_m}) \nonumber \\
   &=& m^{D/y_m}(A_1 + A_2  m^{|y_i|/y_m}) , \nonumber \\
\label{eq:free}
\eea
where $A_1$ and $A_2$ are non-universal constants. 

Fig.~\ref{fig:irfp2} has additional features. The pure gauge theory is confining, and so there is a second
UV fixed point, located at $(g^2,m)=(0,\infty)$.  The presence of a nonzero mass means that
the theory has a mass gap, and so it is easy to conjecture that when the mass becomes large,
the system
``heals over'' to a pure gauge theory, which also has
 a mass gap. Whether there is an additional transition
associated with this behavior is unknown. In addition, many lattice regulated gauge plus fermion theories
are known to confine at strong coupling. If there is end of the critical surface, drawn as the solid
line in Fig.~\ref{fig:irfp2}, there must be an additional transition, but as far as is known, nothing
forces it to have any particular order.

\begin{figure}
\begin{center}
\includegraphics[width=0.5\textwidth,clip]{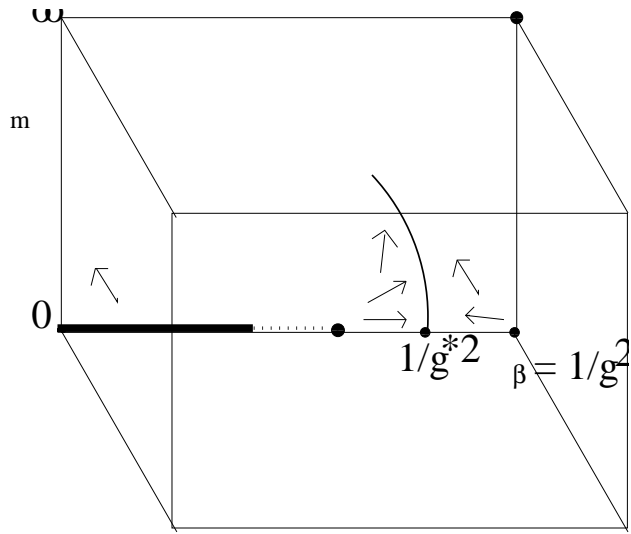}
\end{center}
\caption{ Extending Fig.{\protect{\ref{fig:irfp2}}} by adding an irrelevant direction.}
\label{fig:realirfp}
\end{figure}

Finally, we can add a third direction for all the irrelevant operators present in a lattice calculation.
This produces the horrible Fig.~\ref{fig:realirfp}. I show it to make the point that lattice simulations
are done with a lattice action which of necessity, contains irrelevant operators. We flow onto the
critical surface with enough RG evolution, but getting enough aspect ratio to complete the flow in
an actual  lattice simulation
may not be possible.

Thus, we have a second way of viewing the difference between a confining and a conformal theory,
and the suggestion that if the latter case exists, the phase diagram might be
 more complicated than for a
confining theory. It also introduces a new concept: for an IRFP theory, the gauge coupling
itself is an irrelevant coupling, and the way it affects physics, basically Eq.~\ref{eq:free},
 is quite different
from the way the gauge coupling  in a confining theory does.

Finally, the location of an IR repulsive fixed point on a relevant axis is
physical: one must tune the relevant coupling carefully to approach criticality.
In all other directions its location is scheme dependent: different choices
for blocking transformations move it around.
By design, the IRFP has been tuned to lie someplace on a critical surface (we have
set $m_q=0$), so ``where it is'' is completely scheme dependent.
In simulations, if the relevant quark mass is nonzero,
the correlation length is finite.
 But the variation of the
correlation length (which is a spectral observable) on the gauge coupling is no
different from
the behavior of any physical observable on an irrelevant operator.
What is physical are the exponents, the derivatives of the beta functions
 at the critical point.
For a diagonal set of couplings these are ($\mu$ is a UV scale change)
\bee
\mu \frac{\partial u_i}{\partial \mu} = - y_i(u_i-u_i^*).
\ee
Simple dimensional counting suggests $y_m \sim 1$ for the mass and $y_g\sim 0$
for the gauge coupling.

\section{Practical problems for lattice simulations}
At this point the lattice method for studying a beyond - Standard Model theory seems 
straightforward:
take a QCD code, change the ``3'' of three colors to some other number, write subroutines for
either many flavors of fermions or for fermions in higher representations,
pick a simulation volume, collect data $\dots$. Life is, however, not so simple,
and all simulations done to date are plagued by two kinds of systematic effects. In
 the field they are called ``cutoff effects'' and ``finite size effects.''
 
``Cutoff effects '' describe the problem that the specific lattice action which is simulated
is a bare action defined at the cutoff scale, at a distance of one lattice spacing.
Calculations in quantum field theory are only real predictions when the cutoff has been
removed from the calculation. In lattice QCD simulations, this is referred to
 as ``taking the continuum limit,'' and it is done by tuning the bare $g^2$ to the
 Gaussian fixed point.
The lattice spacing is then traded in for a $\Lambda$ parameter, and dimensionful
predictions can be made in units of $\Lambda$.
Typically, in any lattice version of an asymptotically free theory,
cutoff dependence is strongest when the bare gauge coupling is strong.

``Finite size effects'' arise when  the finite simulation volumes used for numerical
studies force the system to have different behavior than it would in an infinite system.
The classic example of a finite size effect which is physical is when the temporal
length $L_t$ of the lattice is finite. Its length corresponds to a finite temperature $T\propto 1/L_t$.
For a confining theory, a simulation volume which is smaller than the physical confinement scale
will alter the spectrum and also alter the phase of the vacuum. The system size rules
the dynamics.

In a confining theory with a UV attractive fixed point, the lattice spacing shrinks as the
bare gauge coupling is reduced,
and so if the size of the simulation volume is not
increased, one is studying the theory in a box of small physical size.
The squeezed system may have completely different properties from what it would have
in a bigger volume. This is the situation for QCD.

These difficulties merge in the systems which have
been studied to date. The problem  is that, having an IRFP or not, their running coupling runs very slowly.
This occurs because the leading coefficient of the beta function, $b_1$ in Eq.~\ref{2loopbeta2},
is small.
 At one loop, the running coupling is
\bee
\frac{1}{g^2(sL)} = \frac{2b_1}{16\pi^2}\log sL + \rm{constant} .
\label{eq:run}
\ee
For ordinary QCD with two flavors, $b_1=29/3$. Your author has worked
on $N_c=3$ color with $N_f=2$ symmetric representation fermions, where $b_1=13/3$.
This slow running means that for any practical available simulation volume, the only way
 to have a theory which is strongly interacting
at long distance is to make it strongly interacting at short distance,
and thus to risk being contaminated by cutoff effects.

The slow running also means that a small change in the bare coupling can correspond to a
 large change in the physical scale. This can be seen by differentiating Eq.~\ref{eq:run}:
\bee
\Delta \frac{1}{g^2(sL)} = \frac{2b_1}{16\pi^2} \frac{\Delta s}{s}
\ee
For an asymptotically free but slowly walking theory, a tiny change in the bare coupling corresponds to
a greater and greater change of scale.

Of course, if the system has an IRFP, this paragraph makes no sense. The gauge coupling is irrelevant.
Only the amount by which the bare mass is tuned away from zero can affect the correlation length --
asymptotically. As long as the correlation length is finite, tuning $g^2$ will alter it, just
as tuning any irrelevant operator (like an irrelevant coefficient in a lattice action) will
change the correlation length. The problem, of course, is distinguishing 
this ``irrelevant change of length with coupling''
from the true change of length which occurs in a confining and asymptotically free theory.
So we are back to our fundamental problem: how to tell ``no running'' (IRFP) from ``slow running''
(Gaussian fixed point and nothing else) in a noisy lattice Monte Carlo environment with both
a nonzero UV cutoff and a finite simulation volume.

This is very different from ordinary QCD. With its rapidly running coupling,
one can be weakly interacting at short distance and strongly interacting at long distance,
in the same simulation volume.
This can  be observed in (for example) a heavy quark potential behaving as
$V(r) \sim 1/r$ at small $r$ and $\sigma r$ for large $r$.

However, a bug can be a feature.
Slow running is almost no running.
 It is possible to use the finite volume as a diagnostic.
One way to do this is to adapt the familiar finite-size scaling analysis of critical phenomena.
Here the length scale associated with the simulation volume
 $L$ (actually $1/L$) is a relevant operator and  if
the correlation length scales as $\xi \sim (1/m)^{-1/y_m}$ in infinite volume,
then it scales in finite volume proportionally to $L$ times some arbitrary function
of the dimensionless ratio $\xi/L$
\bee
\xi_L = L F(\xi/L) =
 L f(L^{y_m} m_q)   .
\label{eq:fss2}
\ee
Results from many $L$'s can be combined to extract $y_m$. To the particle physicist, the inverse correlation
length could be a mass, or any other dimensionful derived quantity (a decay constant,
for example).

Another way involves using $L$ itself as a scale for a coupling $g^2(L)$
which is derived from the expectation value of some observable.
  Fixed $L$ can be combined with
simulations with fixed boundary conditions to give the ``Schr\"odinger functional''
suite of techniques, which can be used to measure a running coupling and $\gamma_m$
\cite{Luscher:1992an,Luscher:1993gh, Sint:1995ch,
Luscher:1996sc, Sommer:1997xw, Jansen:1998mx, DellaMorte:2004bc}.
I pause to describe this in more detail:

The basic idea is to define a running coupling $g^2(L)$
through the response of a system in a box of size $L$ to its
environment.
In a gauge theory this is done by doing simulations in a finite box
of size $L$ and fixing the value of the spatial
link variables on the faces of the box at Euclidean time $t=0$ and
$t=L$. The boundary conditions involves a free parameter $\eta$.  Call the
resulting partition function the Schr\"odinger functional
\bee Z(\eta) = \int [dU|\exp(-S(U,\eta)).
\ee
 The coupling is then
defined through the variation of the effective action $\Gamma$, the
negative logarithm of the partition function $Z(\eta)$: $\Gamma = -\ln
Z(\eta)$.  In lowest order perturbation theory, $\Gamma$ is equal to
the classical action, which can be computed since the link variables
simply interpolate between their boundary values. At this order the
action is proportional to the inverse squared bare coupling $1/g_0^2$.
The renormalized coupling $g^2(L)$ is defined through
\bee
\frac{\partial \Gamma}{\partial \eta} = \frac{k}{g^2(L)},
\ee
where the constant $k$ is adjusted so that $g^2(L)=g_0^2$ in lowest
order.  For a pure gauge system the quantity $\frac{\partial \Gamma}{\partial \eta}$ is an
expectation value of gauge field variables  on the boundaries.

Next we perform a second simulation in a volume of size $sL_0$
and compute the change in the coupling
\bee
\int_{L_0}^{sL_0} \frac{dL}{L} =  \int_{g^2(L_0)}^{g^2(sL_0)}
\frac{dg^2}{\beta(g^2)} \equiv \int_u^{\sigma(s,u)}\frac{dv}{\beta(v)},
\ee
where the ``step scaling function'', $\sigma[s,u=g^2(L)]=g^2(sL)$,
is the new coupling constant.
The running coupling is found by doing simulations with the same bare
coupling on systems of size $L_0$ and $sL_0$, and, by measuring $u$
and $\sigma(s,u)$, to see how the new coupling depends on the original
one. The ``discrete beta function'' is $\sigma(s,u) -u$.

For asymptotically free theories, one can repeat the matching with various values of $L$,
tuning the bare parameters to stay at fixed $u$ as $L$ is changed. Lattice artifacts
can be removed by comparing the discrete beta function for different values of $L$.
 When the coupling runs quickly enough, one can ``daisy chain''
several  $L_0$ to $sL_0$ pairs to see running over a scale factor $s^n$.
This can be done directly in ordinary QCD, but with beyond Standard Model theories, the coupling
 runs too slowly to do this. Instead, simulations either stop with the discrete beta function
or plot parameterizations of the data.

In another approach \cite{Hasenfratz:2010fi,Hasenfratz:2009kz,Hasenfratz:2009ea},
one can attempt to walk onto the renormalized trajectory by performing simulations
on lattices of various size, blocking the variables, and matching simulations after 
different levels of blocking. Here the schemes basically corresponds to the choice of blocking
transformation.
For example, if a simulation on one lattice size $L_1$ at some $g_1$ and blocked $n_b$ times
with a blocking scale $b$
produces the same observable as a simulation on lattice size $L_2=L_1/b$ does after
$n_b-1$ blocking steps, we would say that the bare couplings must change from $g_1$ to
$g_2$ when the cutoff changes from $a$ to $ba$. Running of bare quantities holding
physical quantities fixed defines a (finite scale) beta function.
This is called the ``Monte Carlo Renormalization Group.''

Essentially all published results for running $g^2$ and  $\gamma_m$ come from one of
these methods. Note that they do not depend on the numerical simulations being done at large
physical volume or that the equilibrium state is 
 in some particular phase.

And there are some explicitly lattice - related  problems. 

The first comes from the choice of lattice fermions \cite{thebook}.
 To make a long story short,
there are three kinds of lattice fermions. The first ones are Wilson or clover fermions.
They explicitly break chiral symmetry due to the presence of lattice operators which prevent
``fermion doubling.'' However, interactions give an additive renormalization to the
bare fermion mass and finding $m_q=0$ (where the quark mass is defined through some 
analog of continuum chiral symmetry, like the PCAC relation, $\partial_\mu A^\mu(x) = 2m_q P(x)$
relating the axial current $A_\mu$, the pseudoscalar current $P$ and the quark mass $m_q$)
involves fine tuning bare lattice quantities.

 The second choice is ``staggered fermions.'' These naturally come in multiples of four
flavors (called ``tastes'' in the literature) and, to make a long story short,
have an exact $U(1)$ chiral symmetry while the expected $SU(4)\otimes SU(4)$ chiral 
symmetry is broken by lattice artifacts. This is called
 the ``flavor problem'' or ``doubling problem.''
 The  conversion of four flavors into an arbitrary number
is well understood in QCD \cite{twostagrefs},
but (as far as I know) only if the theory is chirally broken
are there observables which monitor flavor symmetry breaking, namely the masses
of the would-be Goldstone bosons. They are nondegenerate at any lattice spacing and nonzero
(apart from one pion) at $m_q= 0$.

 Finally, there are chiral fermions, overlap
or domain wall fermions. They are quite expensive and either fail to be
completely chiral or become too expensive to simulate when the gauge field configurations
 become rough
(at strong coupling).

With a small number of flavors, the strong coupling limit of lattice theories with all three
kinds of fermions is confining and chirally broken. There are pseudoscalars whose squared mass
 vanishes with the quark mass, and it is possible to tune bare parameters so that they
vanish. However, with a large number of fermionic degrees of freedom, for example $N_f \sim 7$ flavors
in color $SU(3)$, Wilson fermions develop a first order transition at strong coupling
and $m_q=0$ cannot be reached in a stable ground state: the quark mass jumps
 discontinuously from positive to negative
\cite{Iwasaki:2003de}. In finite volume (where all simulations are done) 
this transition becomes entangled
with the finite temperature confinement - deconfinement transition.

The most unambiguous situation, from the point of view of a simulation, would be to observe
confinement and chiral symmetry breaking at strong and intermediate coupling,  a
beta function which is everywhere negative (so the coupling grows under flow to the IR)
at weak and intermediate coupling, and a region of overlap where both a negative beta function and
confinement are both observed. In this case we have a situation like low-$N_f$ QCD,
where the continuum theory is asymptotically free, confining and chirally broken.

\section{The state of the art -- early autumn 2010}

The cost of $N\times N$ matrix multiplication scales like $N^3$ and so most
 lattice simulations use $N_c=2$ or 3. Referring to Fig.~\ref{fig:ds}, there are then two
ways to approach the conformal window: either by increasing $N_f$ while 
fixing $N_c$ and holding the representation
fixed (to the fundamental, so far), or by fixing $N_f=2$, increasing
 $N_c$ and going to ever higher-dimensional representations. I will call these two choices the
``vertical'' and ''horizontal'' approaches.

There are many simulations of $N_c=2$ and 3 along the vertical axis.
For $N_c=3$ and $N_f \le 8$ the situation seems uncontroversial: The beta function
is negative throughout the range of couplings where it is measured 
\cite{Appelquist:2007hu,Appelquist:2009ty}, and confinement
and chiral symmetry breaking are observed in conventional simulations \cite{severaln8}.
These theories could be technicolor candidates, except that their couplings
apparently run too fast to satisfy phenomenological constraints (they do not ``walk''
and they fail precision electroweak tests).
A recent set of large scale simulations \cite{Appelquist:2009ka}
 has tried to observe the onset of ``condensate enhancement.''
The idea is that, associated with a large $\gamma_m$, the condensate,
expressed dimensionally as $\svev{\bar\psi \psi}/F_\pi^3$, should grow 
compared to the (low-$N_f$)
QCD case. Precision electroweak comparisons have also begun \cite{Appelquist:2010xv}.

Large $N_f$ presents a difficult problem in the chiral limit, because one-loop
corrections to the lowest order formulas for the condensate and $F_\pi$ are proportional
to $N_f$. These one-loop corrections are also the source of finite volume corrections
to chiral observables. Thus large $N_f$ simulations become quite sensitive to the simulation 
volume. This complicates extrapolations to the zero fermion mass limit.

Above $N_f=8$ the situation is ambiguous.
 The original studies of Appelquist,
Fleming and Neil \cite{Appelquist:2007hu,Appelquist:2009ty} saw an IRFP for $N_c=3$,
 $N_f=12$. The problem is that the authors of Refs.~\cite{Fodor:2009wk,Fodor:2009ff}
observe that systems with 8 to 12 flavors of (staggered) fermions appear to be chirally broken
and confining. This is incompatible with the presence of an IRFP unless there
is an additional transition which marks the boundary of
the confining phase with the critical surface of the IRFP.
Perhaps \cite{Deuzeman:2009mh} this transition has been observed. Again the difficulty is
in determining if the transition is induced
 by the finite simulation volume or if it persists in infinite physical volume.

Finally, an $SU(2)$ simulation with 6 flavors of Wilson fermions\cite{Bursa:2010xn}
  claims an IRFP very close
to the first order line, with a large $\gamma_m \sim 0.7$ with a large uncertainty
(presumably because the transition is near the first order line). 

Now for the horizontal branch.
All simulations not using fundamental representation fermions
 use symmetric-representation ones.
 Looking at Fig.~\ref{fig:ds}, we expect that
$N_f=2$ systems are close to conformal, with perhaps larger $N_c$ more likely to
be confining.

 Many groups (a representative list includes
\cite{Catterall:2008qk,Hietanen:2008mr,Hietanen:2009az,Bursa:2009we,DelDebbio:2010hx,DelDebbio:2010hu})
 have studied $N_f=2$ flavors
of adjoint representation fermions in $SU(2)$. All studies to date use Wilson type
quarks.
 The Schr\"odinger functional coupling runs very slowly throughout
the weak coupling phase.  Two groups \cite{Hietanen:2009az,Bursa:2009we}
claim evidence for an IRFP at strong bare gauge coupling.
With Wilson quarks, the
 $m=0$ line in bare coupling space collides with a line of first order
transitions at strong coupling.
The collision point appears to be an IR-repulsive critical point, giving the 
$m=0$ system two UV repulsive fixed points (at $g=0$ and at large $g$).
The IRFP  is found very close to the end of the first
order line. It is unknown whether this compromises the results.

$SU(3)$ with $N_f=2$ symmetric representation fermions is similar. An earlier claim of an IRFP
by   Shamir, Svetitsky and DeGrand \cite{Shamir:2008pb} was not confirmed by a
subsequent simulation by the same authors with a better lattice 
action\cite{DeGrand:2010na}. 
The running coupling just runs very slowly.

Fortunately, the slowly running gauge coupling makes the nearly conformal theory
``conformal for all practical purposes:'' that is, at any value of the bare coupling,
the coupling runs so little that data can be analyzed as if it did not run at all. Then
the mass is the relevant perturbation and all the statistical mechanics machinery for
scaling near a critical point can be employed. This gives $\gamma_m(g^2)$ for $SU(2)$
in a wide variety of ways\cite{Bursa:2009we,DelDebbio:2010hx,DelDebbio:2010hu}.
 It turns out to be small, less than about 0.5,
and consistent with perturbation theory, Eq.~\ref{eq:gmpt}.
 Two measurements in $SU(3)$ with sextet fermions \cite{DeGrand:2009hu,DeGrand:2010na} of
 $\gamma_m$ exploiting this fact gave $\gamma_m<0.6$ over the observed range,
 again perturbative.
Here we also have a calculation of
 a precision electroweak observable \cite{DeGrand:2010tm},
 which
does not look like a phenomenologist's dream for viable technicolor.

\begin{figure}[ht]
\begin{center}
\includegraphics[width=0.8\textwidth,clip]{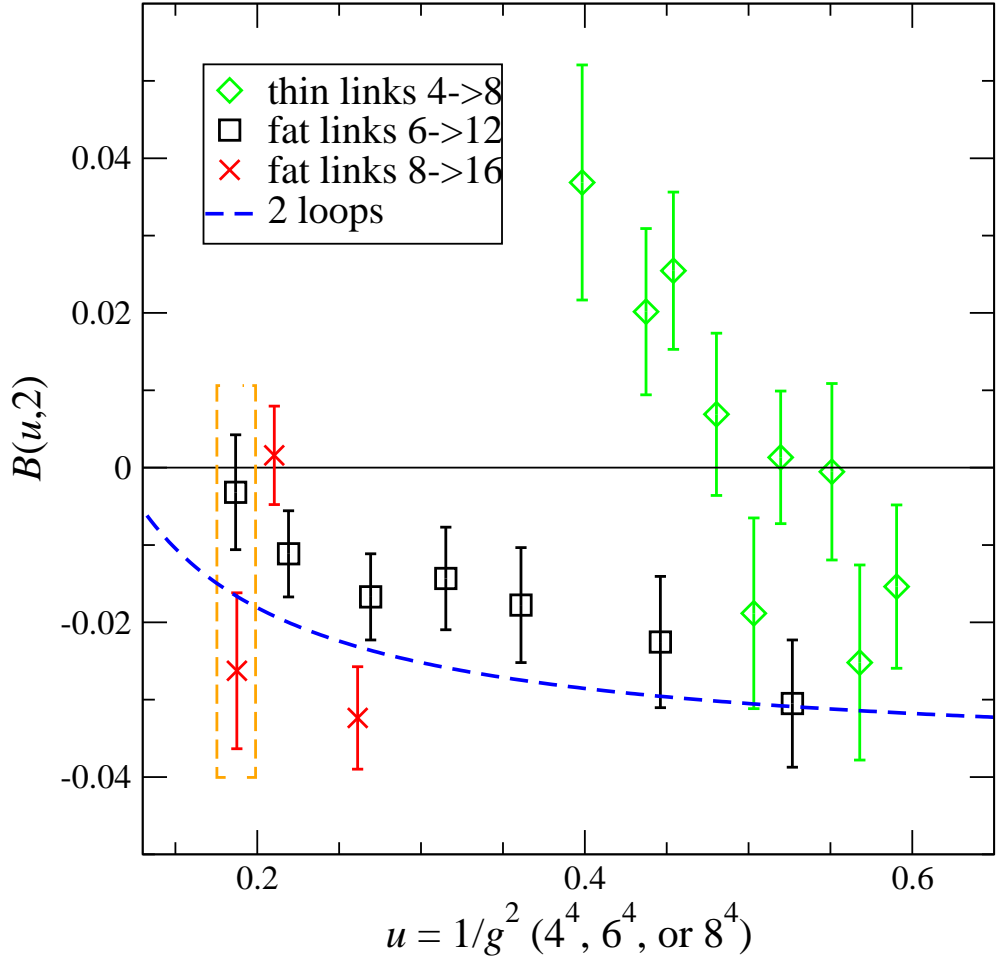}
\caption{Discrete beta function for the scale transformation $L\to2L$,
for the SU(3) gauge theory with sextet fermions. 
``Thin links'' indicates the otherwise unimproved Wilson--clover
 action used in Ref.~\protect{\cite{Shamir:2008pb}};
 ``fat links'' indicates the use of the improved lattice action of
Ref.~\protect{\cite{DeGrand:2010na}}.
The horizontal axis is 
the SF running coupling computed on the smaller lattice, while
the vertical axis gives the change in $1/g^2$ in doubling the lattice
size.
The dashed curve is the two-loop result.  The DBF would change sign at an
infrared-attractive fixed point.
The thin-link data show an IRFP at $g^2\simeq2$, while the fat-link data
exclude this and show no clear evidence of a fixed point before lattice
artifacts set in at $g^2\simeq5$.
\label{fig:Beta}}
\end{center}
\end{figure}
\begin{figure}[ht]
\begin{center}
\includegraphics[width=0.7\textwidth,clip]{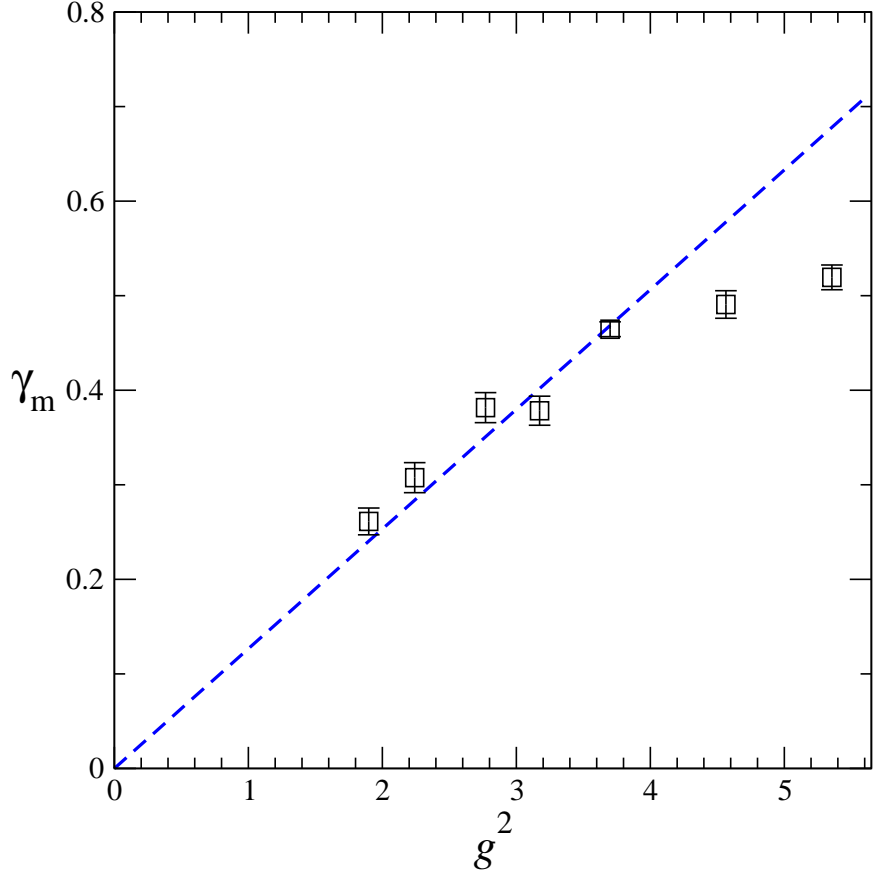}
\caption{Anomalous dimension $\gamma_m$ in the SU(3) gauge theory with sextet
 fermions, determined from Ref.~\protect{\cite{DeGrand:2010na}}.
The dashed line is the one-loop perturbative result.
As the coupling gets stronger the non-perturbative result saturates at values less than 0.6.
\label{fig:gamma}}
\end{center}
\end{figure}
The two cases of difficult running coupling and easy $\gamma_m$ are illustrated by
plots from the author's collaboration's study of $SU(3)$ gauge theory with $N_f=2$ flavors
of sextet fermions. See Figs.~\ref{fig:Beta}  - \ref{fig:gamma}.

\section{Conclusions}
It's a fascinating business, trying to use lattice simulations to search for
and analyze nearly conformal quantum field theories.
People do not agree on what are the important questions, or how to answer them.

The main problem with simulations in this field is that the coupling
 constant runs so slowly
that by the time one forces the theory to be strongly interacting at long distance, it is
strongly interacting at short distance and so not
easily related to a continuum action. Perhaps there is some way to overcome this problem, by 
clever ``lattice action design.'' 
Failing that, it may still be possible to settle the question of whether these near-conformal
theories possess phenomenological viability.
Experiment fortunately puts
stringent constraints on strongly interacting beyond - Standard Model physics.
For example, walking technicolor needs a large $\gamma_m$; if no model has that, then
this approach to a nonperturbative Higgs sector is ruled out.

The last thing I can say: To date, no observed beta function looks like the ``walking
technicolor dream,'' Fig.~\ref{fig:bfnex}b. They all look like simple
deformations of two loop perturbation theory, like
 Fig.~\ref{fig:bfnex}a.

We are still a long way from doing high precision calculations at the level of those
for lattice QCD.
Of course, in QCD we knew what the answer was, before we started: confinement and chiral
symmetry breaking.
Here we don't!


\begin{acknowledgments}
This review is based on a talk at the workshop
``New applications of the renormalization group
method in nuclear, particle and condensed matter physics (INT-10-45W)''
at the Institute for Nuclear Theory at the University of Washington.
I would like to thanks its organizers, M.~Birse, Y.~Meurice, and S-W.~Tsai, for
the opportunity to attend and write this review.
I thank 
T.~Appelquist,
P.~Damgaard, 
L.~Del Debbio,
G.~Fleming,
A.~Hasenfratz,
U.~M.~Heller, 
D.~M.~Kaplan,  
T.~G.~Kovacs,
J.~Kuti,
E.~Neil,
A.~Patella,
 B.~Svetitsky and Y.~Shamir for discussions.
This work was supported in part by the US Department of Energy.

\end{acknowledgments}

\end{document}